\documentclass[aps,prb,amsmath,amssymb,superscriptaddress,preprint]{revtex4-2}

\usepackage{amsmath}
\usepackage{graphicx}
\usepackage{dcolumn}
\usepackage{bm}
\usepackage{upgreek}
\usepackage{natmove}
\usepackage{xcolor}

\begin{document}

\title{Constitutive relations for plasticity of amorphous carbon}

\author{Richard Jana}
\affiliation{Department of Microsystems Engineering, University of Freiburg, Georges-K\"ohler-Allee 103, 79110 Freiburg, Germany}

\author{Julian von Lautz}
\affiliation{Fraunhofer IWM, W\"ohlerstra\ss e 11, 79108 Freiburg, Germany}

\author{S. Mostafa Khosrownejad}

\author{W. Beck Andrews}
\affiliation{Department of Microsystems Engineering, University of Freiburg, Georges-K\"ohler-Allee 103, 79110 Freiburg, Germany}

\author{Michael Moseler}
\affiliation{Fraunhofer IWM, W\"ohlerstra\ss e 11, 79108 Freiburg, Germany}
\affiliation{Cluster of Excellence livMatS, Freiburg Center for Interactive Materials and Bioinspired Technologies, University of Freiburg, Georges-K\"ohler-Allee 105, 79110 Freiburg, Germany}

\author{Lars Pastewka}
\affiliation{Department of Microsystems Engineering, University of Freiburg, Georges-K\"ohler-Allee 103, 79110 Freiburg, Germany}
\affiliation{Cluster of Excellence livMatS, Freiburg Center for Interactive Materials and Bioinspired Technologies, University of Freiburg, Georges-K\"ohler-Allee 105, 79110 Freiburg, Germany}

\begin{abstract}
We deform representative volume elements of amorphous carbon obtained from melt-quenches in molecular dynamics calculations using bond-order and machine learning interatomic potentials. A Drucker-Prager law with a zero-pressure flow stress of $41.2$~GPa and an internal friction coefficient of $0.39$ describes the deviatoric stress during flow as a function of pressure. We identify the mean coordination number as the order parameter describing this flow surface. However, a description of the dynamical relaxation of the quenched samples towards steady-state flow requires an additional order parameter. We suggest an intrinsic strain of the samples as a possible order parameter and present equations for its evolution. Our results provide insights into rehybridization and pressure dependence of friction between coated surfaces as well as routes towards the description of amorphous carbon in macroscale models of deformation.
\end{abstract}


\maketitle

\section{Introduction}

Coatings of amorphous carbon (a-C) are widely used in industrial applications to reduce wear and friction in mechanical contacts~\cite{robertson_diamond-like_2002,erdemir_tribology_2006}. During loading, frictional systems experience severe mechanical conditions that induce subsurface plastic flow.
The resistance of the material to plastic flow can then dominate the frictional response of the system~\cite{rigney_plastic_1979,mishra_analytical_2012,kunze_wear_2014}. For the interpretation of a-C friction experiments, it is therefore important to understand the plastic properties of a-C.

a-C is interesting not just for its wide range of applications but also because it forms an ideal network structure (Fig.~\ref{fig:1}a). Carbon atoms can be sp- (two neighbors), sp$^2$- (three neighbors) or sp$^3$- (four neighbors) hybridized. The pair-distribution function, shown in Fig.~\ref{fig:1}b, vanishes between the first and second neighbor peak. This is in contrast to metallic glasses~\cite{mendelev_using_2007} or even amorphous silicon~\cite{treacy_local_2012} that look more liquid-like~\cite{moras_shear_2018}. It means a-C forms an ideal network; it is the only single-component network-forming glass. 

Since a-C is exclusively produced by means of physical vapor deposition, it only exists in the form of thin films.
Due to the lack of bulk samples, experimental characterization of inelastic mechanical properties has to rely on indentation tests~\cite{oliver_improved_1992,friedmann_thick_1997,kulkarni_nanoindentation_1997,li_fracture_1997,charitidis_nanoindentation_1999,martiinez_study_2001} or the laborious preparation of nanoscale test specimens~\cite{schaufler_determination_2012,kim_mechanical_2014,liu_dual_2017}. Indentation subjects the samples to an inhomogeneous stress field. The extraction of fundamental mechanical properties from indentation is difficult because the inhomogeneity of the stress field must be considered when interpreting indentation experiments.

Within this paper, we use a computational molecular dynamics approach to determine the inelastic properties of a-C. Within computations using representative volume elements, it is straightforward to subject the material to homogeneous deformation. Such molecular dynamics approaches have in the past been used to compute yield of polymer and network glasses. For example, Rottler \& Robbins~\cite{rottler_yield_2001} showed that yield of polymer glasses described by bead-spring models follows a Drucker-Prager~\cite{drucker_soil_1952} or pressure-modified von-Mises law. Their model glasses yielded once the deviatoric (von-Mises) stress $\tau_\textrm{dev}$ exceeded
\begin{equation}
   \tau_\textrm{dev} > \tau_y = \tau_0 + \alpha p,
   \label{eq:DruckerPrager}
\end{equation}
where $p$ is the hydrostatic pressure and $\tau_0$ and $\alpha$ are material properties. 
Similar behavior was found by Moln\'ar et al.~\cite{molnar_densification_2016} for silicate glasses modeled with the BKS potential~\cite{van_beest_force_1990,yuan_local_2001}. Since Eq.~\eqref{eq:DruckerPrager} looks like Amontons' friction law with an adhesive contribution, $\alpha$ is often called the internal friction coefficient. Experimentally, Drucker-Prager-type behavior has been found for polymers~\cite{Bowden1968,Rabinowitz1970}, foams~\cite{deshpande_multi-axial_2001} and metallic glasses~\cite{Davis1975,Lu2003,patnaik_spherical_2004}.

The first objective of this paper is to extract the flow surface of a-C using related methods. We show that an equation like Eq.~\eqref{eq:DruckerPrager} describes the steady-state flow of a-C in our simulations and that this ``flow surface'' does not depend on the initial state of the material. The second objective is to obtain insights into the dynamical approach towards this steady-state flow regime that depends on the initial state of the material. We suggest an empirical relationship describing the evolution of the material with strain. This model is a first step towards a constitutive description of the plastic properties of a-C.

\section{Methods}

\begin{figure}
\centering
    \includegraphics[width=12cm]{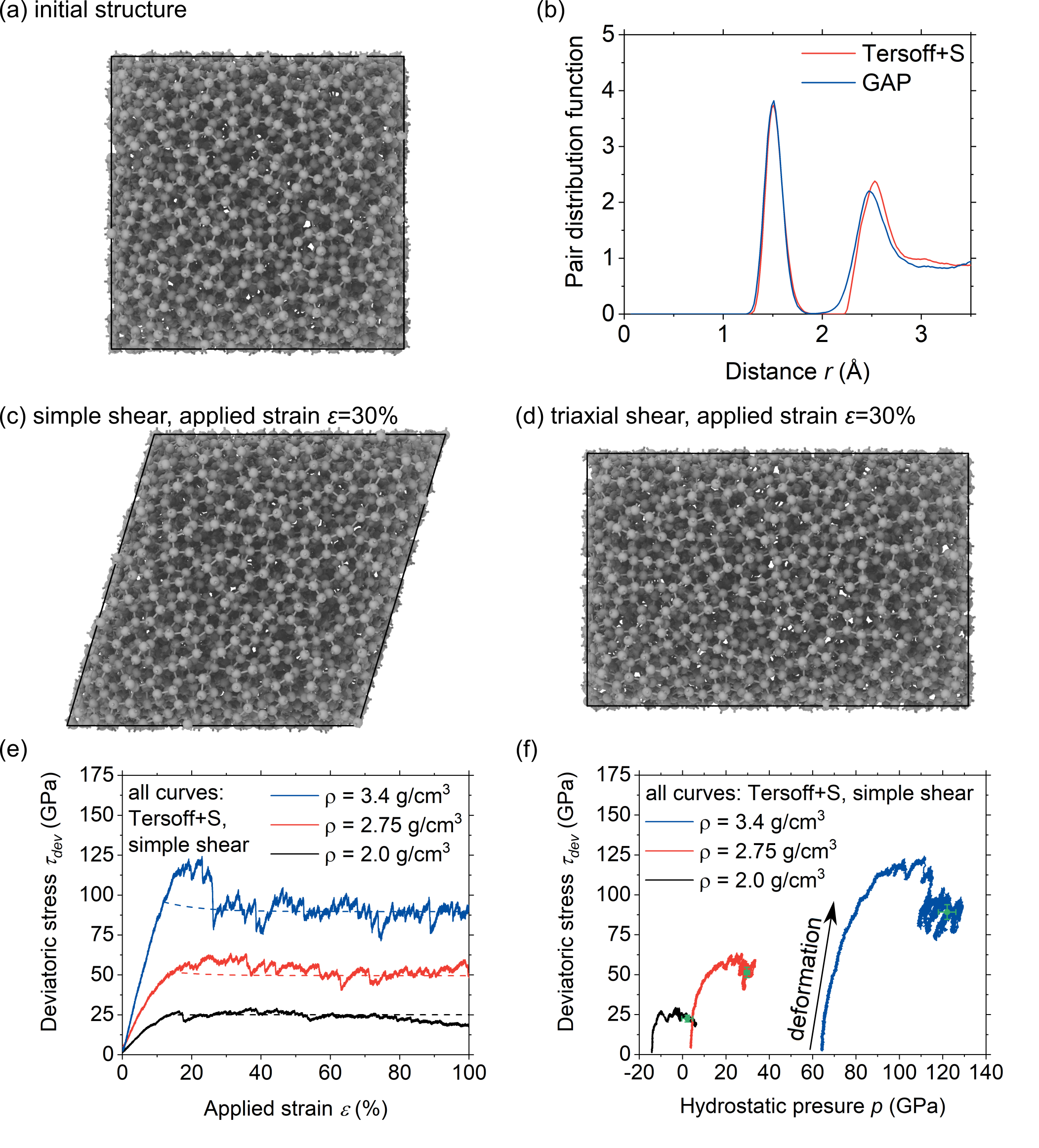}
    \caption{Deformation of amorphous carbon. (a) Initial structure as obtained from a liquid quench. (b) Pair distribution function of the initial structure. We perform molecular dynamics of (c) simple shear and (d) triaxial shear. (e) Examples of stress-strain curves obtained from these calculations (shown here for the Tersoff+S potential). Dashed lines show the solution of the constitutive model (see text). (f) Deviatoric stress a function of hydrostatic pressure used to extract the flow surface. The plot shows the same data as panel (e), but rather than showing $\tau_\textrm{dev}$ as a function of strain $\varepsilon$ we show the hydrostatic pressure $p$ on the $x$-axis. The hydrostatic pressure increases as $\varepsilon$ increases. This is indicated by the arrow marked ``deformation'' that points in the direction of increasing applied strain $\varepsilon$. The flow surface is extracted by averaging $\tau_\textrm{dev}$ and $p$ over the second half of the data. These averages are shown by the green dots in panel (f).}
    \label{fig:1}
\end{figure}

We use two interatomic force models that follow competing philosophies: The screened variant of the Tersoff III potential~\cite{tersoff_modeling_1989,pastewka_screened_2013} (in the following denoted by Tersoff+S) and the Gaussian approximation potential~\cite{bartok_gaussian_2010} (denoted by GAP) as recently parameterized for a-C~\cite{deringer_machine_2017}. The former potential was designed to correctly describe bond-breaking processes~\cite{pastewka_describing_2008} as those continuously occurring during plastic deformation; the latter machine-learning potential gives an accuracy comparable to density-functional theory within the local-density approximation~\cite{martin_electronic_2004} that was used to train it.
Note that the introduction of screening functions~\cite{pastewka_describing_2008} yield a-C properties that are significantly improved over the original Tersoff-III. Both potentials therefore predict structure and mechanical properties within similar uncertainties of experimental measures~\cite{de_tomas_graphitization_2016,de_tomas_transferability_2019,jana_structural_2019} (e.g. see Fig.~\ref{fig:1}b).

Our molecular dynamics calculations start from models of amorphous carbon consisting of $\sim4000$ atoms. These models are obtained by randomly placing the atoms inside a box of given volume. By the choice of volume we create model systems in a range of densities $\rho$ from $2.0$~g~cm$^{-3}$ to $3.5$~g~cm$^{-3}$. All subsequent calculations, liquid quenches and the final deformation, are carried out at this fixed volume. We equilibrate these systems for $25$~ps at $5000$~K after which we quenched the system to $300$~K with time constant $0.5$~ps using a Langevin thermostat. The details of the quench protocol do not appear to matter as the system loses memory of its initial state during plastic deformation. The quench protocol also does not affect structure and elastic properties of the samples, except for very slow quenches where the system may crystallize~\cite{de_tomas_graphitization_2016,de_tomas_transferability_2019,jana_structural_2019}.

We deform these representations of the network glass a-C in direct non-equilibrium molecular dynamics calculations at constant volume. Specifically, we use simple shear (up to $\varepsilon=100\%$ strain, see Fig.~\ref{fig:1}c) and triaxial shear (up to $\varepsilon=50\%$ see Fig.~\ref{fig:1}d) at an applied strain rate of $\dot\varepsilon=10^9$~s$^{-1}$ to map out a representative portion of the flow surface. Shear is imposed by affinely deforming the simulation cell using the deformation gradients
\begin{equation}
  \underline{F}_\textrm{simple} = \left(\begin{array}{ccc}
  1 & \varepsilon & 0 \\
  0 & 1 & 0 \\
  0 & 0 & 1
  \end{array}\right)
\end{equation}
and
\begin{equation}
  \underline{F}_\textrm{triaxial} = \left(\begin{array}{ccc}
  1-\varepsilon & 0 & 0 \\
  0 & \frac{1}{\sqrt{1-\varepsilon}} & 0 \\
  0 & 0 & \frac{1}{\sqrt{1-\varepsilon}}
  \end{array}\right)  
\end{equation}
for simple and triaxial shear, respectively. Note that we refer to $\varepsilon$ as the applied strain throughout this paper. Since $\det \underline{F}=1$, these deformation modes are volume conserving. During deformation, temperature is controlled to $300$~K using a Langevin thermostat with a relaxation time constant of $0.5$~ps. In the case of the simple shear deformation the thermostat was only applied in the direction perpendicular to the shear plane. All simulations are carried out with a time step of $0.5$~fs.

During deformation, we compute the stress tensor using the standard virial expression. Both interatomic potentials yield a glassy disordered carbon network at the quench rates employed here, and their elastic properties are isotropic~\cite{jana_structural_2019}. This isotropic nature of a-C implies that any constitutive equation, such as the yield or flow surface, can only depend
on the principal stresses. From the principal stresses
$\sigma_1$, $\sigma_2$ and $\sigma_3$ in our
simulations, we can calculate the first two invariants of the stress tensor, the hydrostatic pressure
\begin{equation}
    p = \frac{1}{3}\left( \sigma_{1} + \sigma_{2} + \sigma_{3} \right)
\end{equation}
and deviatoric (von-Mises) stress
\begin{equation}
    \tau_{\textrm{dev}} = \sqrt{\frac{1}{2} \Big[ \left( \sigma_{1} - \sigma_{2} \right)^{2} + \left( \sigma_{2} - \sigma_{3} \right)^{2} + \left( \sigma_{3} - \sigma_{1} \right)^{2} \Big] }.
\end{equation}
All our simulations are analyzed in terms of $\tau_\textrm{dev}$ and $p$.

\section{Results and discussion}

Figure~\ref{fig:1}e shows $\tau_\textrm{dev}$ as a function of the applied strain $\varepsilon$ for three select cases: An initially linear (pseudo-) elastic response is followed by yield and then flow of the material at almost constant stress.
The denser samples show shear-softening and we do not find an appreciable difference in the stress-strain response between simple shear and triaxial shear.

Our simulations are carried out at constant volume. We find that during deformation the hydrostatic pressure changes with applied strain. Figure~\ref{fig:1}f shows the deviatoric shear stress $\tau_\textrm{dev}$ as a function of hydrostatic pressure $p$ throughout our simulations. The pressure is constant at small applied strain where the material responds elastically. The nonzero pressure is a residue of the quenching process; we quench at constant volume and do not relax the simulation cell after the quench. The volume elements are under tensile (low density) or compressive (high density) stress. The hydrostatic pressure increases in all cases but then saturates as the material flows. This pressure increases because a-C expands in volume when plastically deformed. Volume expansion has been previously reported in studies of wear of a-C~\cite{kunze_wear_2014} and diamond~\cite{pastewka_anisotropic_2011,moras_shear_2018}. The reason for this expansion in volume is that shearing equilibrates the a-C's structure towards the structure of the liquid phase.\cite{moras_shear_2018}

\begin{figure}
\centering
    \includegraphics[width=12cm]{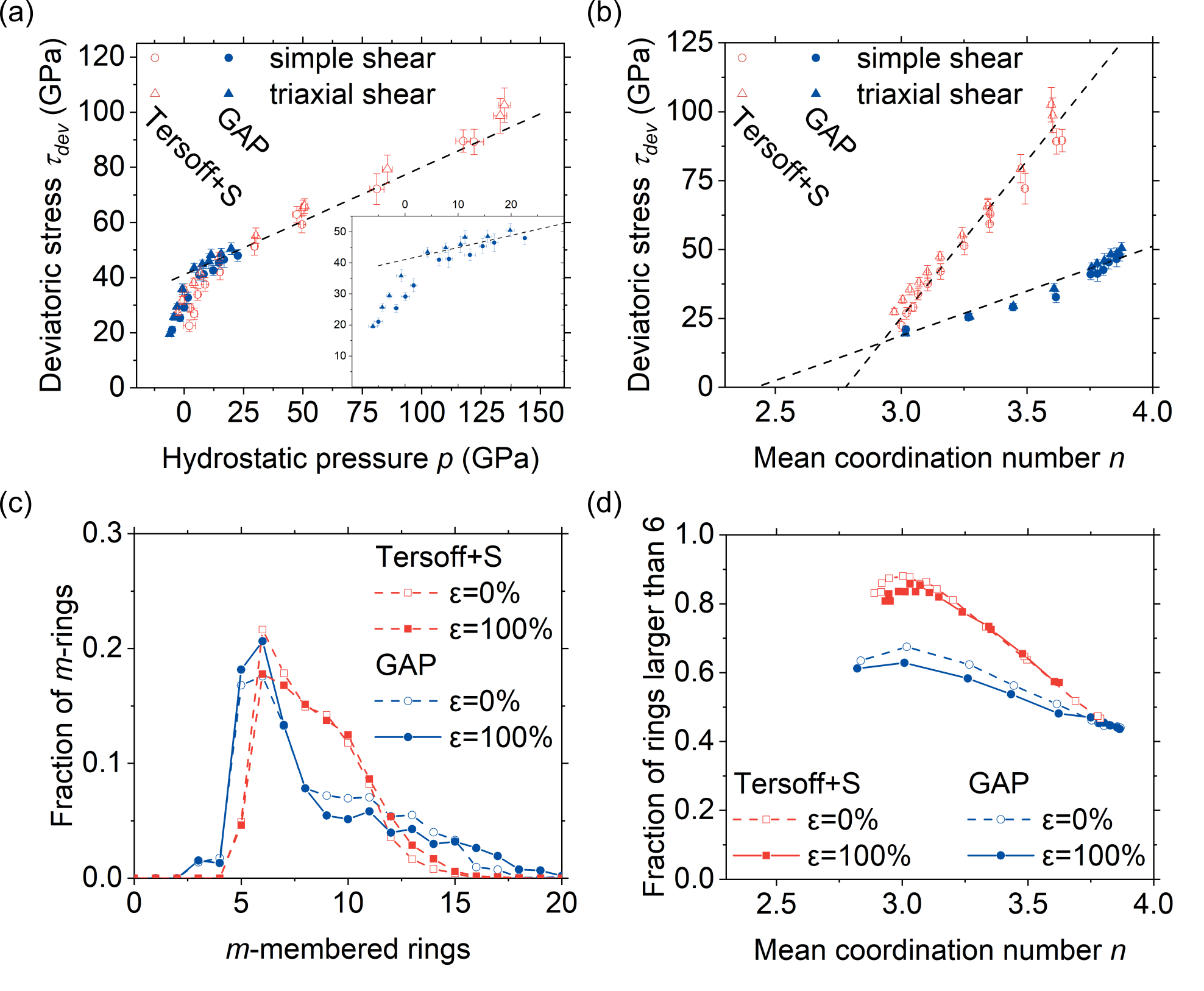}
    \caption{Flow surface of amorphous carbon. (a) Deviatoric stress $\tau_\textrm{dev}$ as function of hydrostatic pressure $p$. The data points are averages over the deformation as shown in Fig.~1f. Error bars are the standard deviation of the fluctuations of $\tau_\textrm{dev}$ and $p$ over the range where they were averaged (see text). The dashed line shows a fit to the Drucker-Prager model, Eq.~\eqref{eq:DruckerPrager}. The inset shows only GAP data points. (b) Deviatoric stress as a function of mean coordination $n$ in the samples. The data is averaged over the same range in applied strain as in panel (a). Dashed lines show linear fits, individually to the Tersoff+S and GAP data. (c) Ring statistics of two systems at $\varepsilon = 0 \%$ and $100 \%$ with $\rho \approx 2.5$ and $2.75$~g/cm$^3$ with GAP and Tersoff+S, respectively. These systems have a mean coordination number of $n \approx 3.25$. The Tersoff+S structure has notably more rings with size between 8 and 11. (d) Fraction of rings with size larger than 6 as a function of the mean coordination number in the structures.}
    \label{fig:2}
\end{figure}

Figure~\ref{fig:1}e and f show only three examples out of a large set of calculations that we have carried out. We varied density (and hence final pressure $p$, cf. Fig.~\ref{fig:1}f), deformation mode (simple shear and triaxial shear), and interatomic force model (Tersoff+S and GAP). For all runs, we average both $\tau_\textrm{dev}$ and $p$ over the final $50\%$ of applied strain for simple shear and over the final $25\%$ of applied strain for triaxial shear. This gives us $\tau_\textrm{dev}(p)$, as shown in Fig.~\ref{fig:2}a. All data collapses onto a single curve, independent of the respective initial condition of our samples and the interaction potential used. At high pressure, there is clearly a linear relationship between $\tau_\textrm{dev}$ and $p$ as described by the Drucker-Prager law, Eq.~\eqref{eq:DruckerPrager}. At $p\approx 0$, $\tau_\textrm{dev}$ drops towards zero, indicating an unjamming transition where the network structure becomes floppy. The inset to Fig.~\ref{fig:2}a shows the behavior of the GAP potential where this drop occurs. Note that an identical drop in shear rigidity at low pressure was found for a fully densified silicate glass~\cite{molnar_densification_2016}.

The dashed line in Fig.~\ref{fig:2}a is a fit to Eq.~\eqref{eq:DruckerPrager} over the portion of the dataset with $p>4.2$~GPa, including data points for both potentials and deformation modes. This yields a parameterization of the flow surface of a-C in terms of the Drucker-Prager law. We obtain $\tau_0=41.2$~GPa and internal friction $\alpha=0.39$. The same universal dependency emerges from two interatomic potentials that were constructed from vastly different philosophies, giving confidence in the robustness of this result.

The Drucker-Prager law constitutes an empirical law for the macroscopic flow of the material. We now turn to the question of whether the resistance to shear (Eq.~\eqref{eq:DruckerPrager}) correlates with a structural measure of the glass. 
The theory of rigidity percolation has identified the mean coordination number $n$ as the central parameter. Mean-field theories~\cite{thorpe_continuous_1983,dohler_topological-dynamical_1980} and numerical calculations of random networks~\cite{he_elastic_1985} predict that random networks loose rigidity for $n<2.4$. The value of $2.4$ is exact for two-dimensional networks and a lower bound for three-dimensional networks. Figure~\ref{fig:2}b shows $\tau_\textrm{dev}$ as a function of $n$, computed by counting neighbors within a cutoff of $r_c=1.85$~\AA{} where the pair distribution has dropped to zero (Fig.~\ref{fig:1}b). We find a linear dependence for both potentials, but with different slopes and different intercepts. Extrapolating $\tau_\textrm{dev}(n)$ to $\tau_\textrm{dev}=0$ we find that the GAP-glass loses rigidity at $n=2.4$, the mean field prediction, while the Tersoff+S-glass loses rigidity at a higher mean coordination of $n \approx 2.8$.

We believe that the difference between the two model glasses relates back to the idea of rigidity percolation. The limit $n=2.4$ only holds for a continuous random network. For general networks, rings with more than $6$ members are floppy and can form floppy regions within the material~\cite{thorpe_continuous_1983}.
In Fig.~\ref{fig:2}c ring statistics~\cite{franzblau_computation_1991} are shown for two systems at $\rho \approx 2.75$~g/cm$^3$. The Tersoff+S structure contains notably more rings with sizes between 8 and 11 and those rings are floppy. Figure~\ref{fig:2}d shows the fraction of rings with sizes larger than six as a function of the mean coordination number $n$ in the structures. At coordination numbers of $3.8$ and above both potentials agree very well, but below Tersoff+S contains a much higher fraction of large, floppy rings towards the coordination where the whole system becomes floppy.

\begin{figure}
\centering
    \includegraphics[width=12cm]{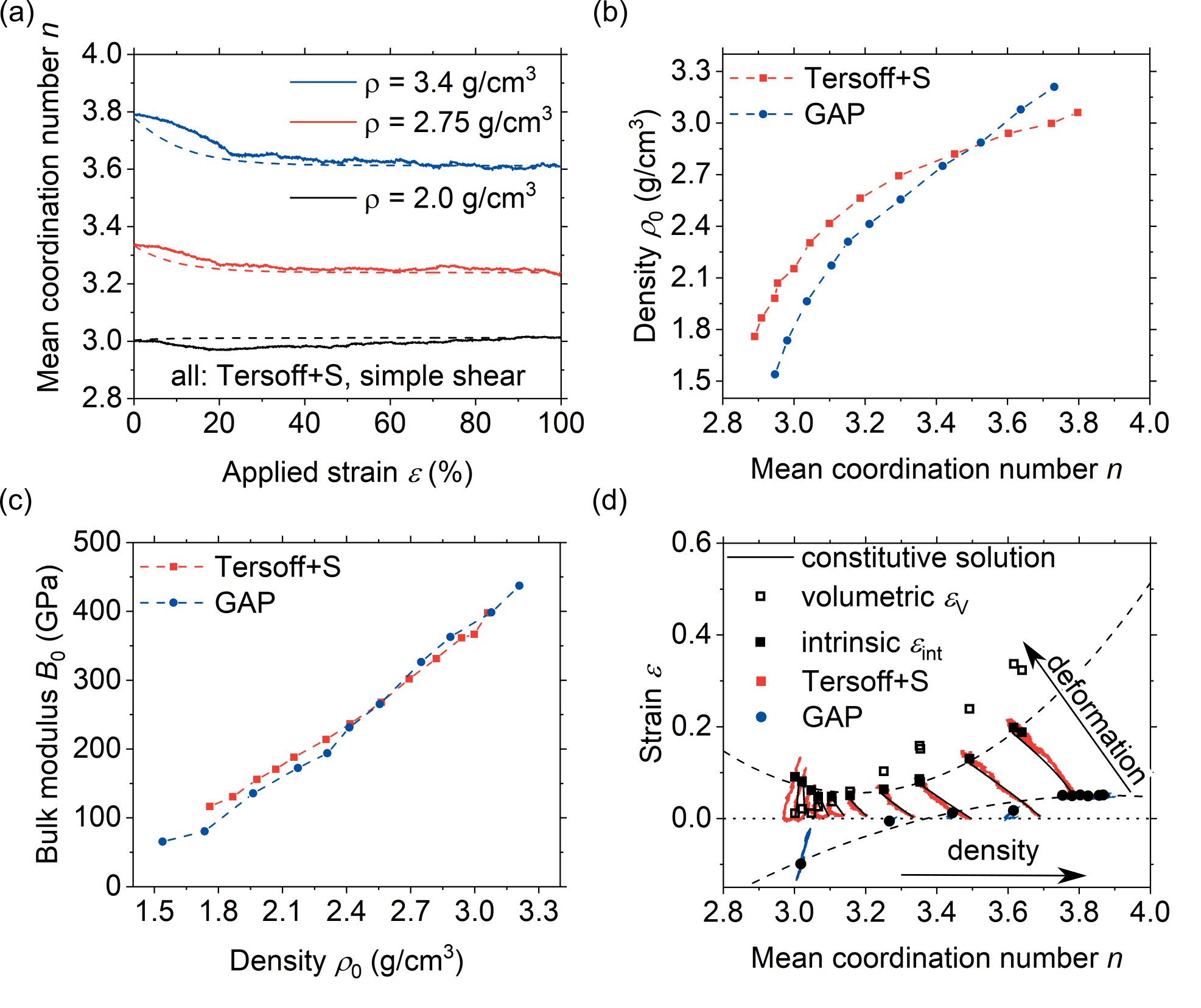}
    \caption{Intrinsic strain. (a) Evolution of the mean coordination number $n$ as a function of applied strain $\varepsilon$ during deformation for the same simulations shown in Fig.~\ref{fig:1}e and f. (Simple shear deformation with the Tersoff+S potential.) Dashed lines show the solution of the constitutive model described in the text. (b) Density $\rho_0$ of the relaxed, stress free structures as a function of coordination number $n$. (c) Bulk modulus $B$ as a function of density $\rho_0$. (d) Evolution of the intrinsic strain $\varepsilon_\textrm{int}$ obtained by subtracting the elastic pressure from the virial pressure obtained throughout the simulation, see Eqs.~\eqref{eq:residual} and \eqref{eq:strain}. Solid black symbols indicate average values for applied strain $\varepsilon = 50$--$100 \%$. Dashed lines are quadratic fits to these average values that show the steady-state $\varepsilon_\textrm{int,0}(n)$. Open symbols show the total volumetric strain $\varepsilon_V$. Solid lines show the solution of the constitutive model. The density of the structures increases from left to right.}
    \label{fig:3}
\end{figure}

In the inelastic regime, our simulations show a drop of the mean coordination number $n$ with applied strain (Fig.~\ref{fig:3}a): the material rehybridizes.\cite{pastewka_running-amorphous_2008,pastewka_atomistic_2010,kunze_wear_2014} Atoms with lower coordination require more volume and hence the pressure during our constant-volume simulations rises. This pressure is partially due to elastic deformation. The relaxed a-C systems follow a unique relationship between density and coordination number, $\rho_0(n)$. Figure~\ref{fig:3}b shows this relationship as obtained from the well-equilibrated simulations reported in Ref.~\citep{jana_structural_2019}. Similarly, the bulk modulus is shown in Fig.~\ref{fig:3}c to uniquely depend on density, $B_0(\rho_0)$. The pressure inside our simulation cell must therefore be given by
\begin{equation}
  p(\rho, n, \varepsilon_{\textrm{int}}) = B_0(\rho_0(n)) \varepsilon_V
  \label{eq:residual}
\end{equation}
with total volumetric strain
\begin{equation}
 \varepsilon_V = \frac{\rho - \rho_0(n)}{\rho} + \varepsilon_{\textrm{int}}.
 \label{eq:strain}
\end{equation}
We call $\varepsilon_{\textrm{int}}$ is the intrinsic (or residual) strain. (Note that in our convention positive volumetric strains are compressions.)

 Figure~\ref{fig:3}d shows the evolution of the intrinsic strain during simple shear deformation at different densities, obtained by solving the generalized equation of state, Eqs.~\eqref{eq:residual} and \eqref{eq:strain}, for $\varepsilon_{\textrm{int}}$. The figure also shows average values over the strain range of $\varepsilon = 50$--$100 \%$ (solid symbols) and empirical quadratic fits to these values.
For Tersoff+S, the trajectories start at $\varepsilon_{\textrm{int}} \approx 0$, showing that our structures are initially free of intrinsic strain but that it builds up during deformation. Only the curves for the highest density structures start at $\varepsilon_\textrm{int}\approx 0.05$. With applied strain, the mean coordination number $n$ decreases and $\varepsilon_{\textrm{int}}$ increases. The GAP trajectories also start at $\varepsilon_\textrm{int} \approx 0$, but $n$ and $\varepsilon_{\textrm{int}}$ show less variation with applied strain than the Tersoff+S trajectories. The average values for $\varepsilon_{\textrm{int}}$ are lower than for Tersoff+S.

From the total volumetric strain $\varepsilon_V$ (open symbols in Fig.~\ref{fig:3}d), we see that the intrinsic strain is the dominant contribution to the overall volumetric strain in the system. The evolution of the hydrostatic pressure in our simulations can therefore be related to the evolution of the intrinsic strain during deformation. Our interpretation of the intrinsic strain is that deformation leads to a distortion of the atomic structure that changes its volume. This distortion may be difficult to quantify in geometric terms, similar to the difficulty of finding geometric order parameters that can distinguish between a-Cs quenched at different rates. (See Ref.~\citep{jana_structural_2019} for a detailed discussion.)

The coordination number $n$ alone is therefore not a sufficient order parameter for the description of the state of the material. A constitutive model for a-C  requires the introduction of an additional state variable, for example the intrinsic strain $\varepsilon_\textrm{int}$ directly. As shown in Fig.~\ref{fig:3}d, the combined macroscopic state vector $(n,\varepsilon_\textrm{int})$ evolves towards a manifold of steady-state values that is shown by the dashed line in Fig.~\ref{fig:3}d and can be described by a functional relationship $\varepsilon_\textrm{int,0}(n)$. The relaxation towards this steady-state behavior is for example shown in Fig.~\ref{fig:1}e and \ref{fig:3}a. Assuming rate-independence with a characteristic relaxation strain $\varepsilon_c$, an approximate evolution law for the state vector in the spirit of a relaxation time approximation is
\begin{equation}
    \frac{d \varepsilon_\textrm{int}}{d \varepsilon} = -\frac{\varepsilon_\textrm{int} - \varepsilon_\textrm{int,0}(n_0)}{\varepsilon_c}
    \quad\textrm{and}\quad
    \frac{d n}{d \varepsilon} = -\frac{n - n_0}{\varepsilon_c}.
    \label{eq:relax}
\end{equation}
The target of the relaxation, the steady-state coordination number $n_0(\varepsilon_\textrm{int}, n)$ depends on the current state, as can be directly seen in Fig.~\ref{fig:3}d. We can extract the steady-state behavior by following along the pathway of deformation in Fig.~\ref{fig:3}d. Given $\delta n(\varepsilon_\textrm{int})=dn/d\varepsilon_\textrm{int}$ as the slope of the evolution of $n(\varepsilon_\textrm{int})$ in this figure, we find $n_0$ as the solution of the nonlinear equation
\begin{equation}
    n_0 - n = \delta n(\varepsilon_\textrm{int}) \left[\varepsilon_\textrm{int,0}(n_0) - \varepsilon_\textrm{int}\right]
    \label{eq:nonlinear}
\end{equation}
for each state $(n,\varepsilon_\textrm{int})$.
Ingredients to this constitutive law are the tangent $\delta n(\varepsilon_\textrm{int})$, the steady-state intrinsic strain $\varepsilon_\textrm{int,0}(n_0)$ and the relaxation constant $\varepsilon_c$. Note that the relaxation constants for $\varepsilon_\textrm{int}$ and $n$ in Eq.~\eqref{eq:relax} could differ and would need to be determined from additional calculations not presented here.

The solid lines in Fig.~\ref{fig:3}d show a solution of this model for $\varepsilon_c=0.1$ within the order parameter space. As shown by the dashed lines in Fig.~\ref{fig:3}a, this solution describes the evolution of the coordination number $n$ with applied strain $\varepsilon$ well. It also serves as a partial explanation for the shear-softening behavior seen at high densities. The deviatoric stress $\tau_\textrm{dev}$ drops (see Fig.~\ref{fig:1}a) because the coordination number decreases and this weakens the material. Using the linear $\tau_\textrm{dev}(n)$ dependency shown in Fig.~\ref{fig:2}b, we obtain the dashed lines in Fig.~\ref{fig:1}e that qualitatively capture the response of the material.

Note that the set of equation presented here cannot describe the response to a change in the density that occurs along the dashed line in Fig.~\ref{fig:3}d. To describe this behavior, Eq.~\ref{eq:nonlinear} must couple to the density $\rho$ or the total pressure $p$, and additional calculations are required to extract an approximate mathematical description of this coupling required for a fully-formulated constitutive law.

Finally, we note that there are large differences between the behavior of the Tersoff+S and the GAP glass: Tersoff+S has a stronger tendency towards rehybridization. This means that for GAP, we cannot extract $\delta n(\varepsilon_\textrm{int})$ as $n$ shows variation only by $0.05$ and the resolution with which we can resolve changes in $n$ depends on the total number of atoms in our unit cell. We expect that a similar picture emerges for GAP but are at present limited to $\sim 4000$ atoms because of the computational cost of the GAP potential. Despite these differences in the structural changes of the material, the flow surface (Fig.~\ref{fig:2}a) appears independent of the choice of interatomic potential.

\section{Summary and conclusion}

In summary, we find that steady-state flow of a-C is described by a Drucker-Prager law. Model glasses obtained from two different interatomic potentials collapse onto the same Drucker-Prager law, giving confidence to the extracted parameters. Our model glasses behave differently with regards to the evolution of the mean coordination number of the system (or alternatively, the  numbers of sp$^3$-, sp$^2$- and sp-hybridized atoms). We can extract a constitutive relationship for these models that involves an intrinsic strain of these structures as an additional order parameter. These results are the first parameterization of the flow surface of a-C. They have relevance for understanding the rehybridization and friction of a-C surfaces in sliding contact that has been observed experimentally~\cite{erdemir_tribology_2006,pastewka_running-amorphous_2008} and in simulations~\cite{pastewka_atomistic_2010,kunze_wear_2014}. Our results also open a route for the development of constitutive models for macroscale calculations of plastic deformation or fracture in a-C.

\section*{Acknowledgments}

We thank Gabor Cs\'anyi and Volker Deringer for providing an early version of the a-C GAP and Jan Grie\ss er for useful discussion. This research was supported by the Deutsche Forschungsgemeinschaft (DFG grant PA 2023/2) and the European Research Council (ERC-StG-757343). All molecular dynamics calculations were carried out with LAMMPS~\cite{plimpton_fast_1995}. ASE~\cite{hjorth_larsen_atomic_2017} and OVITO~\cite{stukowski_visualization_2010} was used for pre-processing, post-processing and visualization. Computations were carried out on NEMO (University of Freiburg, DFG grant INST 39/963-1 FUGG) and JURECA (J\"ulich Supercomputing Center, project ``hfr13'').

\providecommand{\newblock}{}

\end{document}